\documentclass[a4paper,11pt]{article}
\usepackage{pos}
\usepackage{xcolor,graphicx,soul}

\newcommand{\lsim}
{\;\raisebox{-.3em}{$\stackrel{\displaystyle <}{\sim}$}\;}
\newcommand{\gsim}
{\;\raisebox{-.3em}{$\stackrel{\displaystyle >}{\sim}$}\;}

\newcommand\al{\alpha}
\newcommand\be{\beta}
\newcommand\tb{\tan\beta}

\newcommand\CBA{c_{\beta - \alpha}}

\newcommand\ReDiag{\mathop{%
  \raise .5pt\hbox{[}%
  \widetilde{\mathrm{Re}}%
  \raise .5pt\hbox{]}}}
\newcommand\ReOffDiag{\mathop{%
  \raise .5pt\hbox{$\llbracket$}%
  \widetilde{\mathrm{Re}}%
  \raise .5pt\hbox{$\rrbracket$}}}

\newcommand\Mh{M_h}
\newcommand\MH{M_H}
\newcommand\MA{M_A}
\newcommand\MHp{M_{H^\pm}}

\newcommand\refeqs[1]{Eqs.~(\ref{#1})}

\newcommand\refse[1]{Sect.~\ref{#1}}

\newcommand\citere[1]{Ref.~\cite{#1}}
\newcommand\citeres[1]{Refs.~\cite{#1}}

\newcommand{\CP}{{\cal CP}}
\newcommand{\cp}{{\CP}}

\newcommand{\tev}{\,\, \mathrm{TeV}}
\newcommand{\gev}{\,\, \mathrm{GeV}}

\newcommand\fb{\ensuremath{\mbox{fb}}}
\newcommand\ab{\ensuremath{\mbox{ab}}}

\newcommand\ifb{\ensuremath{\fb^{-1}}}
\newcommand\iab{\ensuremath{\ab^{-1}}}

\newcommand{\De}{\Delta}

\newcommand{\sig}{\sigma}

\def\reffi#1{\mbox{Fig.~\ref{#1}}}

\def\Ga{\Gamma}
\def\ga{\gamma}
\def\de{\delta}
\newcommand{\la}{\ensuremath{\lambda}}
\newcommand{\ka}{\ensuremath{\kappa}}
\newcommand\kala{\ensuremath{\kappa_{\lambda}}}
\newcommand\laSM{\ensuremath{\lambda_{\mathrm{SM}}}}
\newcommand{\lahhh}{\ensuremath{\la_{hhh}}}
\newcommand{\lahhH}{\ensuremath{\la_{hhH}}}

\newcommand{\hotf}{\ensuremath{h_{125}}}
\newcommand{\hnf}{\ensuremath{h_{96}}}

\definecolor{Orange}{named}{orange}
\definecolor{Purple}{named}{purple}
\definecolor{Lightblue}{cmyk}{0.9,0.1,0.1,0.3}
\definecolor{dgelborange}{cmyk}{0.,0.3,0.5, 0.}
\definecolor{Lila}{rgb}{0.5,0.,1}
\definecolor{Darkgreen}{rgb}{0.,.7,0.2}

\graphicspath{{figs/}}

\title{Future Perspectives for Higgs Physics}

\author*[a]{Sven Heinemeyer}

\affiliation[a]{Instituto de F\'isica Te\'orica (UAM/CSIC),\\
Universidad Aut\'onoma de Madrid, 
Cantoblanco, 28049, Madrid, Spain}

\emailAdd{Sven.Heinemeyer@cern.ch}

\abstract{Future perspectives for Higgs physics are outlined. First it is
shown that the discovered Higgs boson cannot be the Standard Model (SM)
Higgs boson, motivating the investigations of Higgs sectors beyond
the SM (BSM). The secure future, the HL-LHC, and the agreed upon future,
an $e^+e^-$ collider, are briefly discussed. The importance of theory
calculations for the full exploitation of the (anticipated) experimental
data is emphasized. Finally, the complementarity of collider based Higgs
physics and other types of experiments, such as gravitational wave
observatories is briefly outlined.}

\FullConference{%
  41st International Conference on High Energy physics - ICHEP2022\\
  6-13 July, 2022\\
  Bologna, Italy \hfill IFT--UAM/CSIC--22-143
}

\begin{document}
\maketitle


\section{Why the Higgs is not the SM Higgs}
\label{sec:intro}

The discovery of a Higgs boson at ATLAS and CMS in 2012, nearly 50 years
after its prediction, was a
milestone in high-energy physics~\cite{Aad:2012tfa,Chatrchyan:2012xdj}.
Within theoretical and experimental uncertainties this new particle is
consistent with the existence of a Standard-Model~(SM) Higgs boson at a mass
of~$\sim 125 \gev$~\cite{Khachatryan:2016vau}. 
While no sign of Beyond Standard-Model (BSM) physics was (yet)
discovered at the LHC, the measurements of Higgs-boson production and
decay rates, which are known 
experimentally to a precision of roughly $\sim 10-20\%$, leave
ample room for BSM interpretations.
Consequently, one of the main tasks of the current and future
LHC runs, as well as experiments beyond the LHC is to determine whether 
this particle, \hotf, forms part of the Higgs sector of an extended
model.

From the theory side this situation is clearer. There are unambiguous
data that tell us that the SM cannot be the ultimate theory. The most
compelling ones are:
\begin{enumerate}

\item gravity is not included in the SM\\[-1.5em]
  
\item the hierarchy problem (the stability of the Higgs-boson mass
  w.r.t.\ quantum corrections)\\[-1.5em]
  
\item no unification of the three forces in the SM\\[-1.5em]
  
\item Dark Matter is not included in the SM\\[-1.5em]
  
\item the Baryon Asymmetry of the Universe (BAU) cannot be explained in
  the SM\\[-1.5em] 

\item neutrino masses are not included in the SM\\[-1.5em]

\item some experimental data are not described correctly within the SM\\
  ($(g-2)_\mu$, flavor anomalies, \ldots)\\[-1.5em]

\end{enumerate}
This clearly indicates that we have to extend the SM, and thus that the
Higgs boson discovered at the LHC cannot be the SM Higgs boson. However,
the really important question is\\[.5em]
{\large\bf Q:} does the BSM physics that undoubtly
exists has any (relevant) impact on the Higgs sector?\\[.5em]
The two obvious ways to answer this question are\\[.5em]
{\large\bf A\boldmath{$_1$:}}
measure the characteristics of the \hotf\ with highest precision,\\
{\large\bf A\boldmath{$_2$:}}
search for additional Higgs bosons ({\em above} and
{\em below} $125 \gev$),\\[.5em]
together with a possibly less obvious way\\[.5em]
{\large\bf A\boldmath{$_3$:}}
test ``other predictions'' of the BSM Higgs sector(s).\\[.5em]
In these proceedings we will briefly discuss these future perspectives
for Higgs-boson physics.


\section{The secure future: HL-LHC}
\label{sec:hl-lhc}

The high luminosity stage of the LHC, the HL-LHC, is approved and
expected to collect up to $3\,\iab$, each in ATLAS and in CMS,
i.e.\ about 20 times more integrated luminosity than collected so
far. This will allow to measure the properties of the \hotf\ with a much
higher precision as compared to the current accuracy, as well as to
extend substiantially the reach for new BSM Higgs bosons. Here we just
list three promient examples in these directions.

The \ka\ framework~\cite{LHCHiggsCrossSectionWorkingGroup:2013rie} was
devised after the Higgs-boson discovery to test the experimental data on
Higgs-boson rate measuerments at the LHC for deviations from the
SM.%
\footnote{Over the last $\sim 10$~years more refined methods have been
  developed, in particular the SMEFT framework, see e.g.\
  \citeres{LHCHiggsCrossSectionWorkingGroup:2016ypw,Passarino:2016pzb} and
  references therein.}%
~Effectively, each $\ka_p$ parametrizes multiplicativly the deviation of
the coupling of the particle~$p$ with the Higgs-boson. The expected
deviations of the Higgs couplings from the SM limit depend on the mass
scale of the new physics.
In general Two Higgs Doublet Model (2HDM)-type models (including the
case of the Minimal Supersymmetric Standard Model (MSSM)) one
expects deviations from the SM predictions of roughly the
following size for the different couplings (here for Yukawa
type~II)~\cite{Baer:2013cma},
\begin{align}
\kappa_V \approx 1 - 0.3 \% \left(\frac{200\gev}{\MA}\right)^4, \;
\kappa_t = \kappa_c \approx 1 - 1.7 \% \left(\frac{200\gev}{\MA}\right)^2, \;
\kappa_b = \kappa_\tau \approx 1 + 40 \% \left(\frac{200\gev}{\MA}\right)^2 .
\label{kappa-target-susy}
\end{align}
Here $\ka_{V,t,c,b,\tau}$ denotes the scale factors for the couplings to
massive SM gauge bosons, top, charm and bottom quarks and to $\tau$~leptons,
respectively, and $\MA$ denotes the BSM Higgs-boson mass scale.
In composite Higgs models one typically expects effects of~\cite{Dawson:2013bba}
\begin{align}
\kappa_V \approx 1 - 3 \% \left(\frac{1\tev}{f}\right)^2, \quad
\kappa_F \approx 1 - (3\mbox{--}9) \% \left(\frac{1\tev}{f}\right)^2 ,
\label{kappa-target-comp}
\end{align}
where $\kappa_F$ is a generic scale factor for the couplings to all
fermions, and $f$ is the compositeness scale. It can be seen that,
depending on the scale of new physics (in this case $\MA$ or $f$)
deviations in the sub-percent range for the couplings to gauge-bosons
can be expected. For the couplings to fermions, depending on the type
and the concrete model, deviations in the per-cent range could be realized.

\begin{figure}[htb!]
\begin{center}
  \includegraphics[width=0.95\textwidth,height=15em]{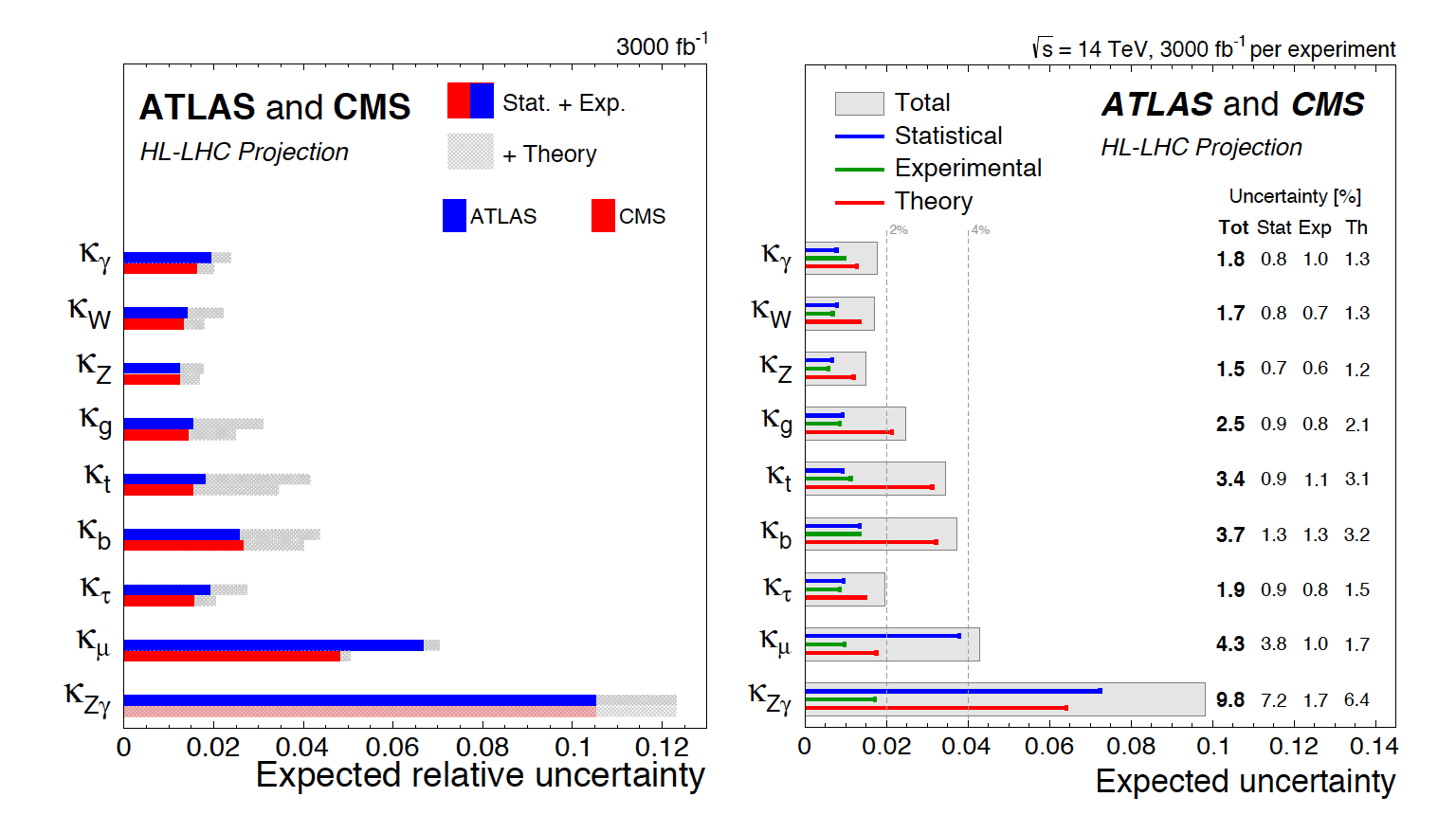}
  \caption{Anticipated improvements at the HL-LHC in the precision of
    $\ka$~measurements~\cite{Cepeda:2019klc}.
  Left: expected uncertainty for ATLAS and CMS; right: combined
  uncertainty (see text).} 
\label{fig:HL-LHC:kappa}
\end{center}
\vspace{-2em}
\end{figure}

In \reffi{fig:HL-LHC:kappa} the accuracies of various $\ka$'s expected
to be reached at the HL-LHC are shown~\cite{Cepeda:2019klc}. The left
plot depicts the relative uncertainties for ATLAS (blue/gray) and CMS
(red/gray), where blue/red indicates the combined experimental
uncertainty, and gray shows the inclusion of theory uncertainties. The
right plot demonstrates the ATLAS/CMS combined result with the
statistical (systematical) uncertainties indicated in blue (green), the
theory uncertainties in red, and the combined result in gray. Here it
should be kept in mind that in order to extract absolute values in the
$\ka$~framework at the (HL-)LHC some theory assumptions must be
made. From \reffi{fig:HL-LHC:kappa} it can be seen that many $\ka$'s
will be known at the $\sim 2\%$ level, with $\sim 4\%$ uncertainties for
$\ka_{t,b,\mu}$ and $\sim 10\%$ uncertainty for $\ka_{Z\ga}$. While this
constitutes an important improvement over the current result, it may
not be sufficient to reach the sensitivity to new physics scales as
indicated in \refeqs{kappa-target-susy} and (\ref{kappa-target-comp}).

\begin{figure}[htb!]
\begin{center}
  \includegraphics[width=0.95\textwidth]{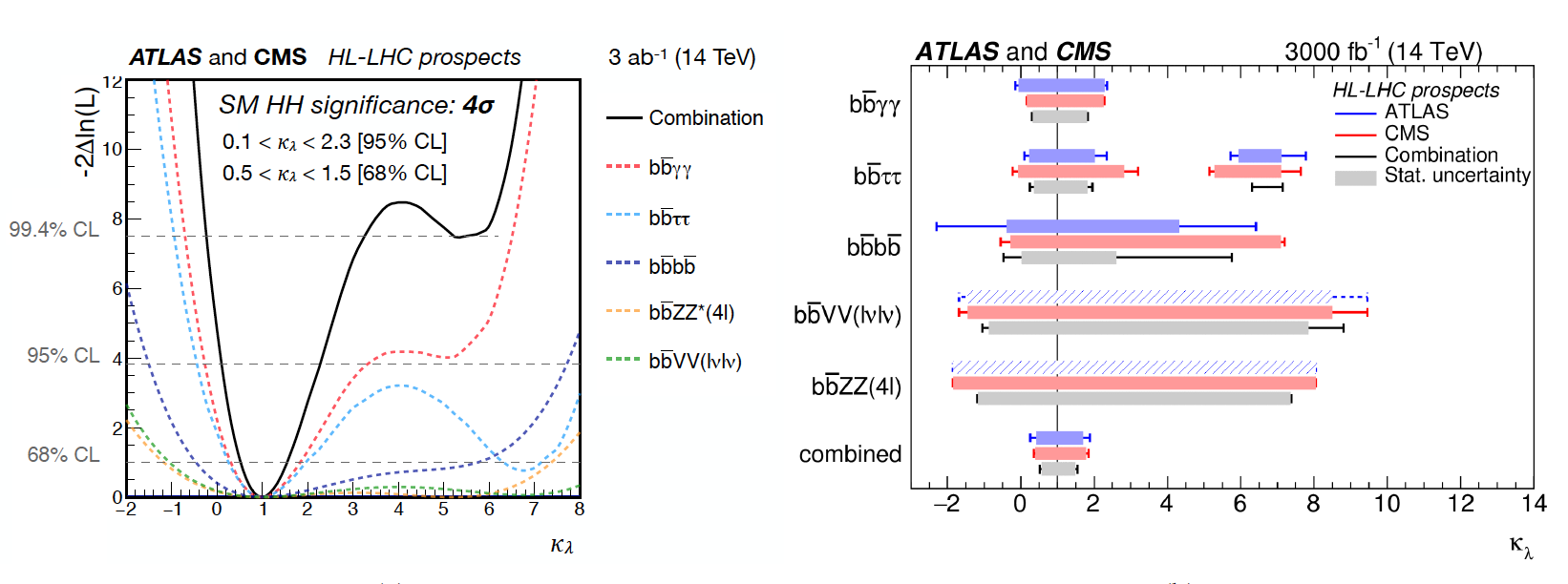}
  \caption{Anticipated improvements at the HL-LHC in the precision of
    $\kala$~measurements~\cite{Cepeda:2019klc} (see text).}
\label{fig:HL-LHC:kala}
\end{center}
\vspace{-2em}
\end{figure}

In \reffi{fig:HL-LHC:kala} the expected relative precision for the SM Higgs
self-coupling, $\kala := \lahhh/\laSM$ is
shown~\cite{Cepeda:2019klc}. The right plot demonstrates the precisions
for the various final states analyzed (ATLAS in blue, CMS in red, the
combination in black, and the pure statistical uncertainty in gray). The
left plot shows the $\chi^2$ results for the various final states (in
color) and for the combined result in black. At the HL-LHC a $\sim 50\%$
precision for \kala\ can be expected (if the SM value $\kala = 1$ is
realized). While this shows at the $\sim 2\,\sig$ level that $\kala \neq 0$,
it does not correspond to a precision measurement of this quantity.

\begin{figure}[htb!]
\begin{center}
  \includegraphics[width=0.75\textwidth]{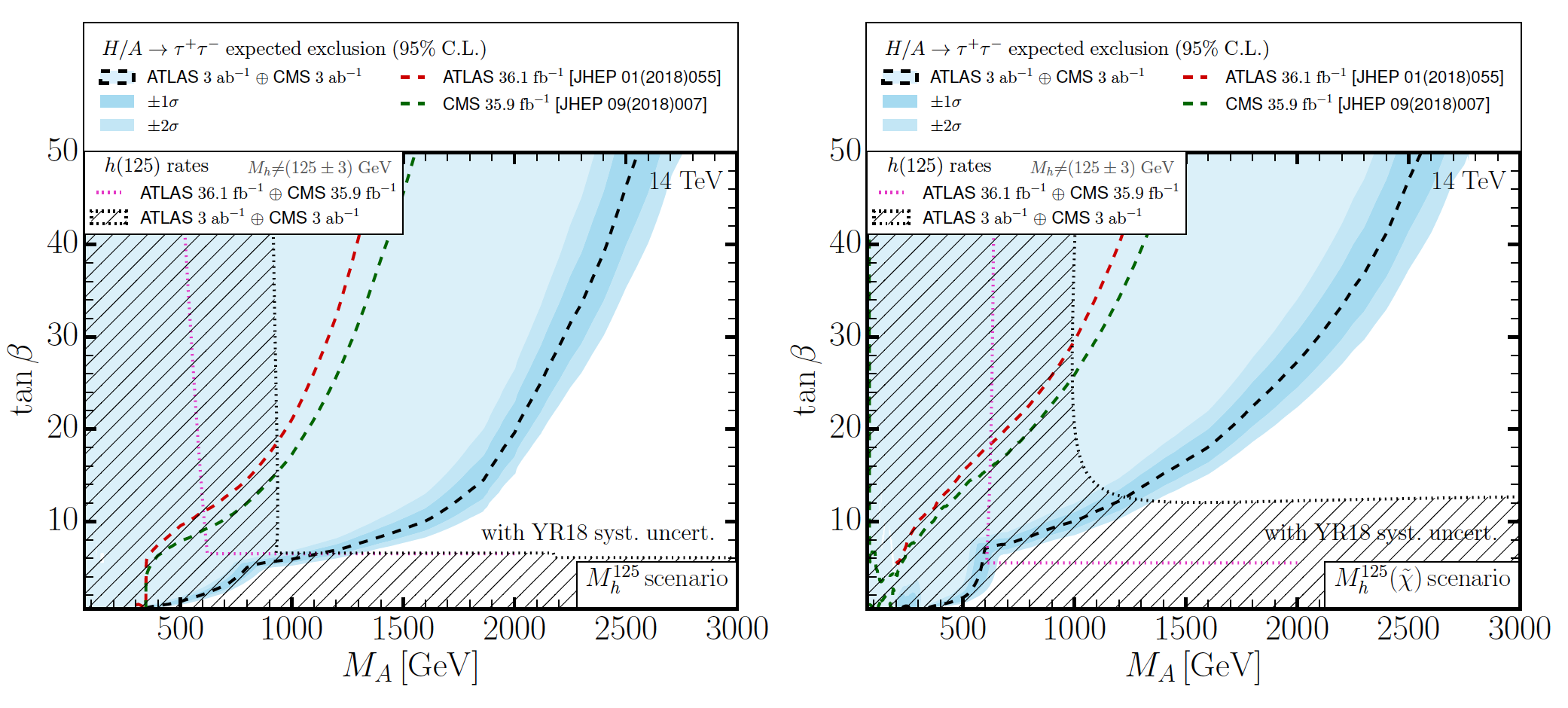}
  \caption{Anticipated reach of the HL-LHC in the search for heavy MSSM
    Higgs bosons~\cite{Bahl:2020kwe} (see text).}
\label{fig:HL-LHC:HAtautau}
\end{center}
\vspace{-2em}
\end{figure}

In \reffi{fig:HL-LHC:HAtautau} we turn to the anticipated
projections for the HL-LHC reach for heavy MSSM Higgs
bosons in the $\tau\tau$ decay channel~\cite{Bahl:2020kwe} in the
$\MA$-$\tb$ plane (where $\MA$ is the 
$\cp$-odd Higgs boson mass and $\tb := v_2/v_1$ is the ratio of the two
vacuum expectation values). The left (right) plot shows the results in
the $M_h^{125}$ ($M_h^{125}(\tilde\chi)$) benchmark scenario, see
\citere{Bagnaschi:2018ofa} for details. The (then) current limits are
shown in red (green) dashed for ATLAS (CMS), based on the first year
Run~2 data. The pink dotted line indicates the limits from \hotf\ rate
measurements. The HL-LHC expectations are shown in black dashed for the
Higgs boson searches and in black dotted for the \hotf\ measurements. In
both scenarios the BSM Higgs-boson mass scale can be tested up to
$\sim 1.2 \tev$. On the other hand, values of $\MA \gsim 2.5 \tev$
remain completely out of reach.

\begin{figure}[htb!]
\vspace{-0.5em}
\begin{center}
  \includegraphics[width=0.52\textwidth]{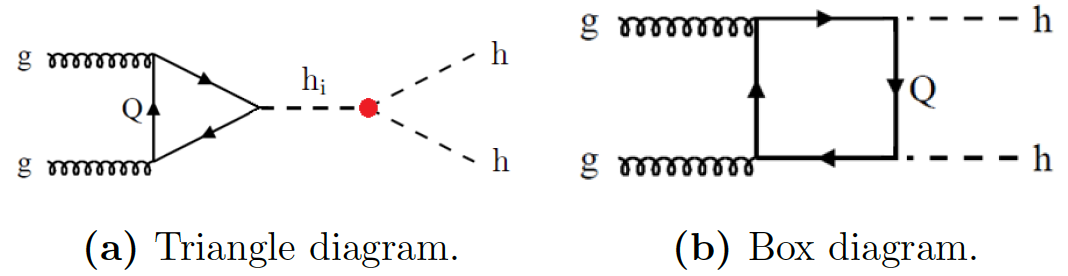}\\
  \includegraphics[width=0.72\textwidth]{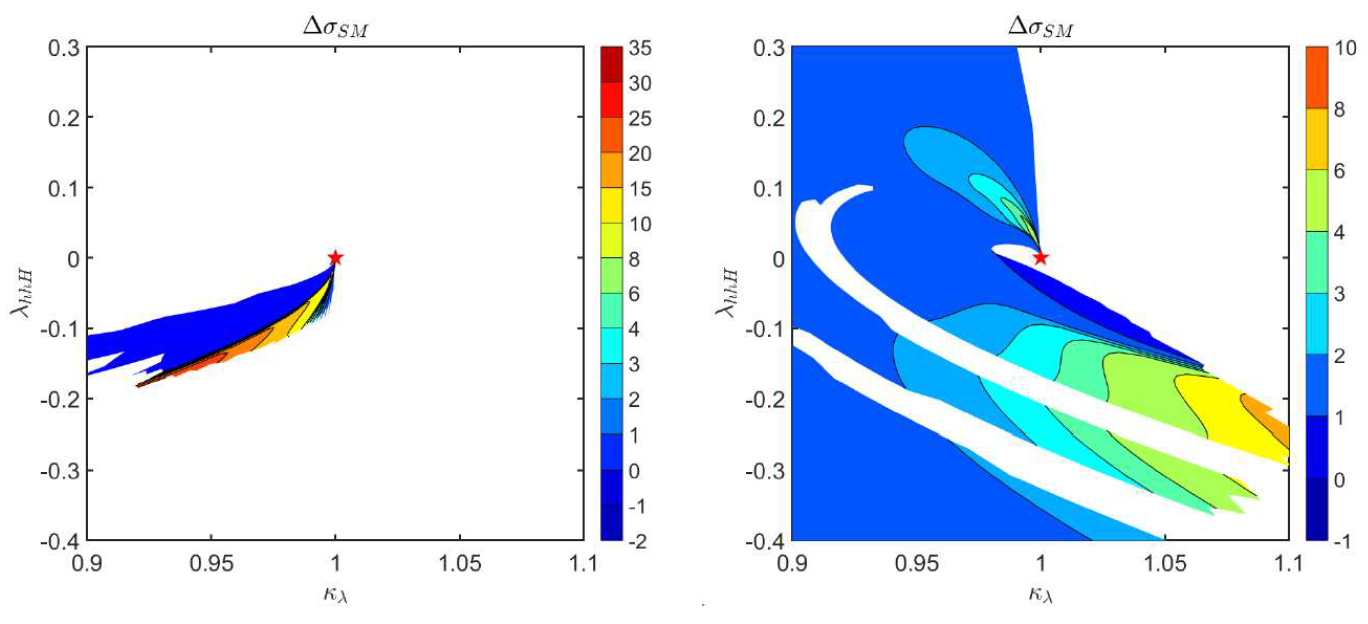}
  \caption{Upper part: Feynman diagrams contributing to $gg \to hh$ in
    the 2HDM ($h_i = h, H$).
    Lower part: di-Higgs production cross section in a selected
    benchmark plane (see text) in terms of
    $\De\sig_{\rm SM} \sim 38\,\fb/4.5$ (see text), taken from
    \citere{Kateryna-tfm}.} 
\label{fig:HL-LHC:lahhH}
\end{center}
\vspace{-1.5em}
\end{figure}

Finally, in \reffi{fig:HL-LHC:lahhH}~\cite{Kateryna-tfm} we demonstrate
that the HL-LHC may also have access to triple Higgs couplings (THCs)
that involve BSM Higgs bosons. In \citere{Kateryna-tfm} the di-Higgs
production cross section $gg \to hh$ is evaluated in the 2HDM using an
accordingly modified version of the code
{\tt HPAIR}~\cite{Abouabid:2021yvw,Grober:2017gut}, where the $h$ is the
\hotf. The upper line of
\reffi{fig:HL-LHC:lahhH} shows the contributing diagrams. The left
diagram depicts contributions with a top triangle and $h,H$ in the
$s$-channel, where $H$ is the heavy $\cp$-even Higgs boson. These
diagrams involve the THCs $\lahhh$ and $\lahhH$, respectively. The right
diagram depicts the (continuum) top-box contribution. The lower row of
\reffi{fig:HL-LHC:lahhH} presents the results for the di-Higgs cross
section in a 2HDM Yukawa type~I benchmark plane~\cite{Arco:2020ucn}
(with $\tb = 10$, all heavy Higgs-boson masses are equal, and
$m_{12}^2 = \MH^2 \cos^2\al/\tb$, with $\cos(\be-\al)$ and $\MH$ as free
parameters). 
The results are projected into the \kala-\lahhH\ plane for
$\CBA$ negative (positive) in the left (right) plot. The color coding
indicates by how many standard devitions, $\De\sig_{\rm SM}$, the 2HDM
cross section differs from the SM cross section. For $\De\sig_{\rm SM}$
the precision expected for $\kala = 1$ has been used,
$\De\sig_{\rm SM} = 38\,\fb/4.5$ (with
$\sig(gg \to hh)_{\rm SM} = 38\,\fb$ at NLO, which can be ``seen'' at
the $4.5\,\sig$ level). The color coding indicates that a non-zero value
of \lahhH, but $\kala \approx 1$ can lead to a sizable enhancement of
the cross section in the 2HDM via the resonant enhancement of the
$H$-exchange diagram. This indicates that the HL-LHC may also have
access to BSM THCs (see \citere{Kateryna-tfm} for details).


\section{The agreed upon future: \boldmath{$e^+e^-$} collider(s)}
\label{sec:e+e-}

In the high-energy physics community there is a large consensus that the
next large experiment following the LHC should be an $e^+e^-$
collider~\cite{ee-consensus}. As an example, the final text of the
European Strategy for Particle Physics Update~\cite{esppu} states:
``\ldots there are compelling scientific arguments for a new
electron-positron collider operating as a 'Higgs factory'. Such a
collider would produce copious Higgs bosons in a very clean environment,
[and] would make dramatic progress in mapping the diverse interactions of the
Higgs boson with other particles \ldots'' The four most advanced
proposals are
the ``International Linear Collider'' (ILC)~\cite{ilc-web},
the ``Compact Linear Collider'' (CLIC)~\cite{clic-web},
the ``Future Circular ($e^+e^-$) Collider'' (FCC-ee)~\cite{fcc-ee-web},
and the ``Circular $e^+e^-$ Collider'') (CEPC)~\cite{cepc-web}. The
projected final energy stages are $1 \tev$, $3 \tev$, $365 \gev$ and
$360 \gev$, respectively. 

\begin{figure}[htb!]
\begin{center}
  \includegraphics[height=6em]{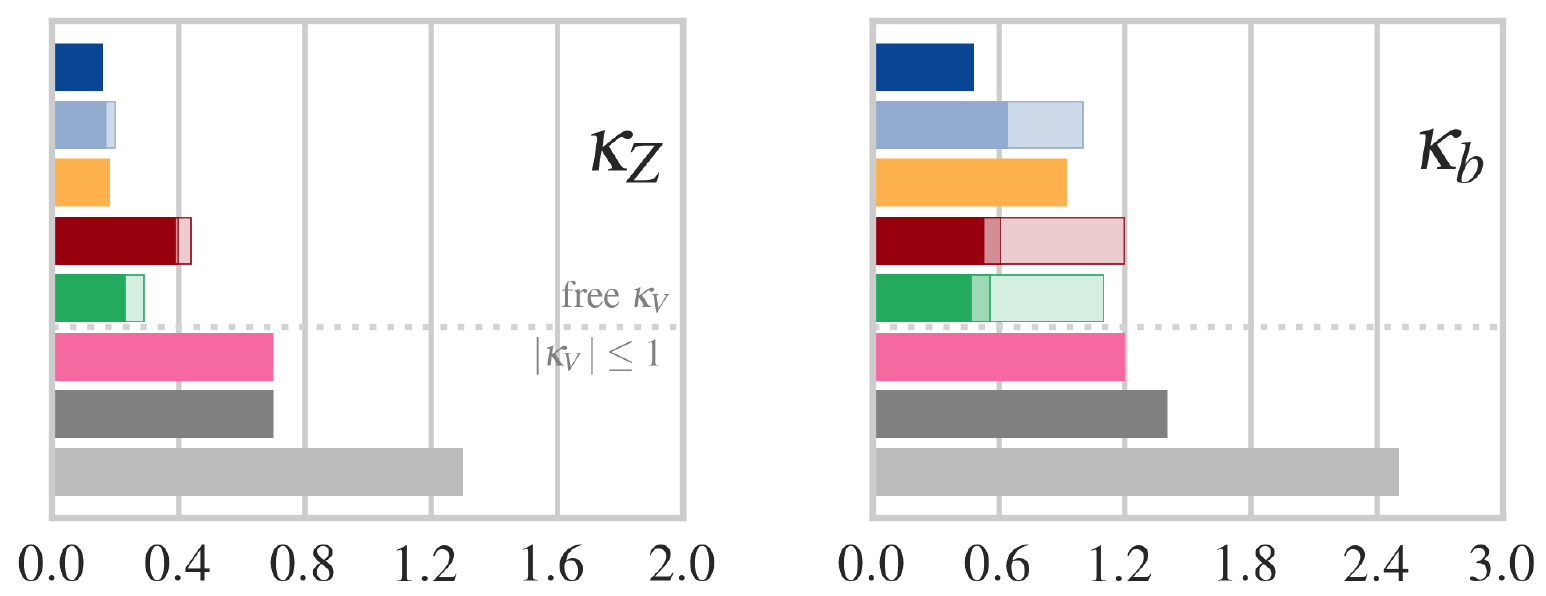}
  \includegraphics[height=3.5em]{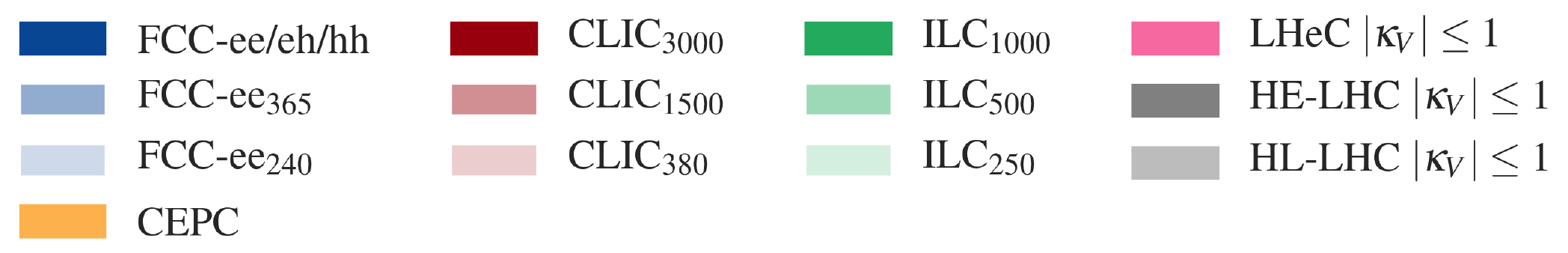}
  \caption{Projected precision in $\ka_Z$ and $\ka_b$ for various future
    colliders~\cite{deBlas:2019rxi}.} 
\label{fig:kappa-comp}
\end{center}
\vspace{-2em}
\end{figure}

\begin{figure}[htb!]
\begin{center}
  \includegraphics[width=0.60\textwidth]{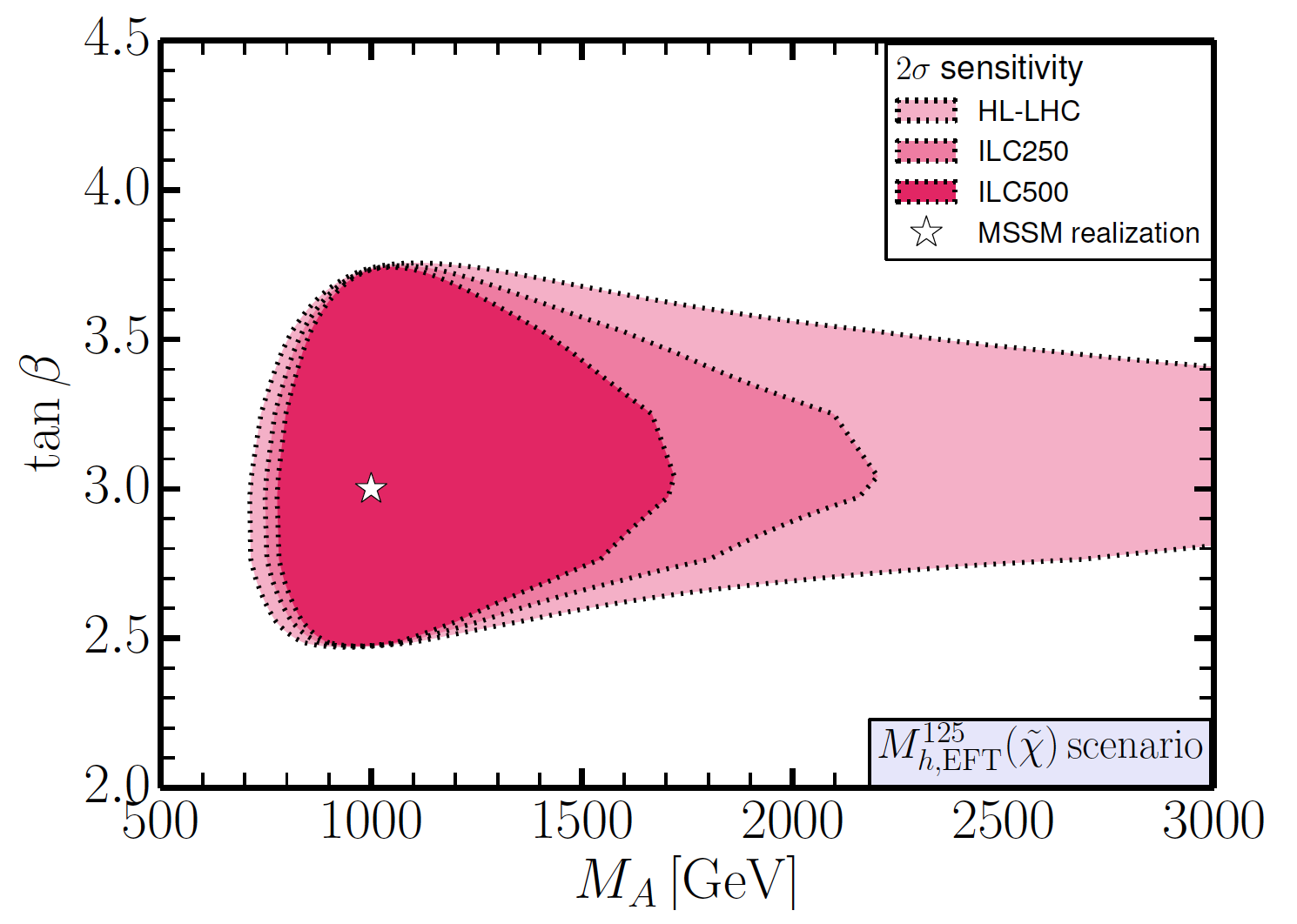}
  \caption{Indirect $2\,\sig$ constraints in the $\MA$-$\tb$ plane of
    the $M_{h,{\rm EFT}}^{125}(\tilde\chi)$ scenario
   from prospective Higgs-boson signal-rate measurements at the HL-LHC
   and the ILC~\cite{Bahl:2020kwe} (see text).}
\label{fig:ILC-MA-limit}
\end{center}
\vspace{-2em}
\end{figure}

In \reffi{fig:kappa-comp}~\cite{deBlas:2019rxi} we review the anticipated
precision of $\ka_Z$ and $\ka_b$ for various future collider options,
where the ILC (CLIC, FCC-ee, CEPC) projections are shown in green
(burgundy, blue, beige). One can observe that w.r.t.\ the HL-LHC
projections (gray, where $|\ka_V| < 1$ had to be assumed for a
converging fit) a strong improvement can be observed. All four $e^+e^-$
collider options very roughly yield the same level of precision. Only with this
kind of improvement the precision required by \refeqs{kappa-target-susy}
and (\ref{kappa-target-comp}) can be reached. As a concrete example we
show in \reffi{fig:ILC-MA-limit} the indirect $2\,\sig$ constraints in
the $\MA$-$\tb$ plane of the $M_{h,{\rm EFT}}^{125}(\tilde\chi)$
MSSM benchmark scenario
from prospective \hotf\ signal-rate measurements at the HL-LHC
and the ILC~\cite{Bahl:2020kwe}, where the point $\MA = 1 \tev$ and
$\tb = 3$ had been assumed to be realized. The light pink area shows the
constraints 
using the projected HL-LHC precision in the \hotf\ couplings. It
can be observed that no upper limit on $\MA$ can be set. This changes
once the ILC precision is assumed with the ILC250 (ILC500) ranges shown
in medium (dark) pink. In this case an upper limit on the so far
unobserved new Higgs-boson mass scale of $\sim 2100 \gev$ ($\sim 1600 \gev$)
can be obtained. This sets a clear target for direct searches at future
colliders. 

\begin{figure}[htb!]
\begin{center}
  \includegraphics[width=0.55\textwidth]{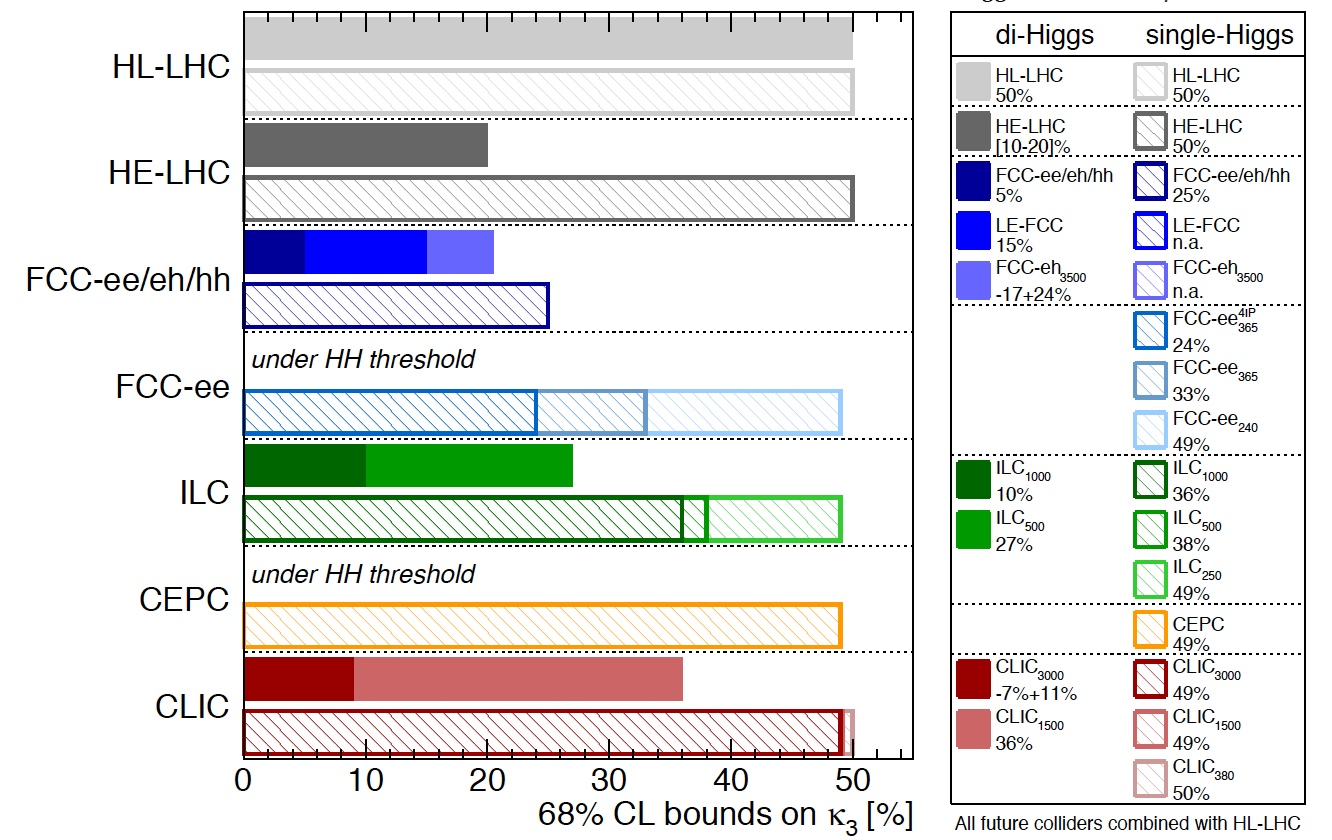}
  \caption{Sensitivity at the 68\% CL on \kala\ at various future
    colliders~\cite{deBlas:2019rxi}.}
\label{fig:kala-comp}
\end{center}
\vspace{-2em}
\end{figure}

Finally, in \reffi{fig:kala-comp} we show the projections for the
precision at the 68\% CL on $\kala$ ($\kappa_3$ in the plot) obtainable
at various future collider options~\cite{deBlas:2019rxi}. The light bars
indicate the preicision using single Higgs production, where
\lahhh\ enters only at the loop level, a method on which circular
$e^+e^-$ colliders must rely, as their center-of-mass energy does not
reach the di-Higgs production threshold. The dark colors indicate the
precision based on di-Higgs production, which can be obtained at the
linear collider options. One can observe that the determination via
di-Higgs productions leads to substantially better sensitivities, going
down to the level of $\sim 10\%$ at the ILC1000 and CLIC3000, a factor
of $\sim 5$ improvement w.r.t.\ the HL-LHC expectations. This will allow
to probe the Higgs potential in the SM, but also in BSM models (for
the 2HDM see, e.g., \citere{Arco:2021bvf}, where also an analysis of the
sensitivity to \lahhH\ can be found). Here it should be kept in mind
that the results in \reffi{fig:kala-comp} assume $\kala = 1$, and the
accuracies can change substantially for other values, see below.

A second main task of a future $e^+e^-$ collider, besides the precision
measurement of the \hotf, is the search for new Higgs bosons, which
would clearly indicate a BSM Higgs sector. In models which extend the SM
by doublets and singlets, the production of two BSM Higgs bosons
together is favored over the single production. Owing to other
theoretical and experimental constraints, this limits the discovery
reach of an $e^+e^-$ collider to roughly $\MH \lsim \sqrt{s}/2$ (where
$\MH$ represents generically the BSM Higgs mass scale). This
substantially limits the discovery reach of circular machines for masses
above $125 \gev$, whereas linear machines, depending on their final
energy stage, have a unique discovery potential. The situation is
different for BSM Higgs bosons below $125 \gev$. Here all future
$e^+e^-$ collider proposals possess an important coverage for a
discovery and/or (subsequent) precision measurements.

\begin{figure}[htb!]
\vspace{-0.5em}
\begin{center}
  \includegraphics[width=0.45\textwidth]{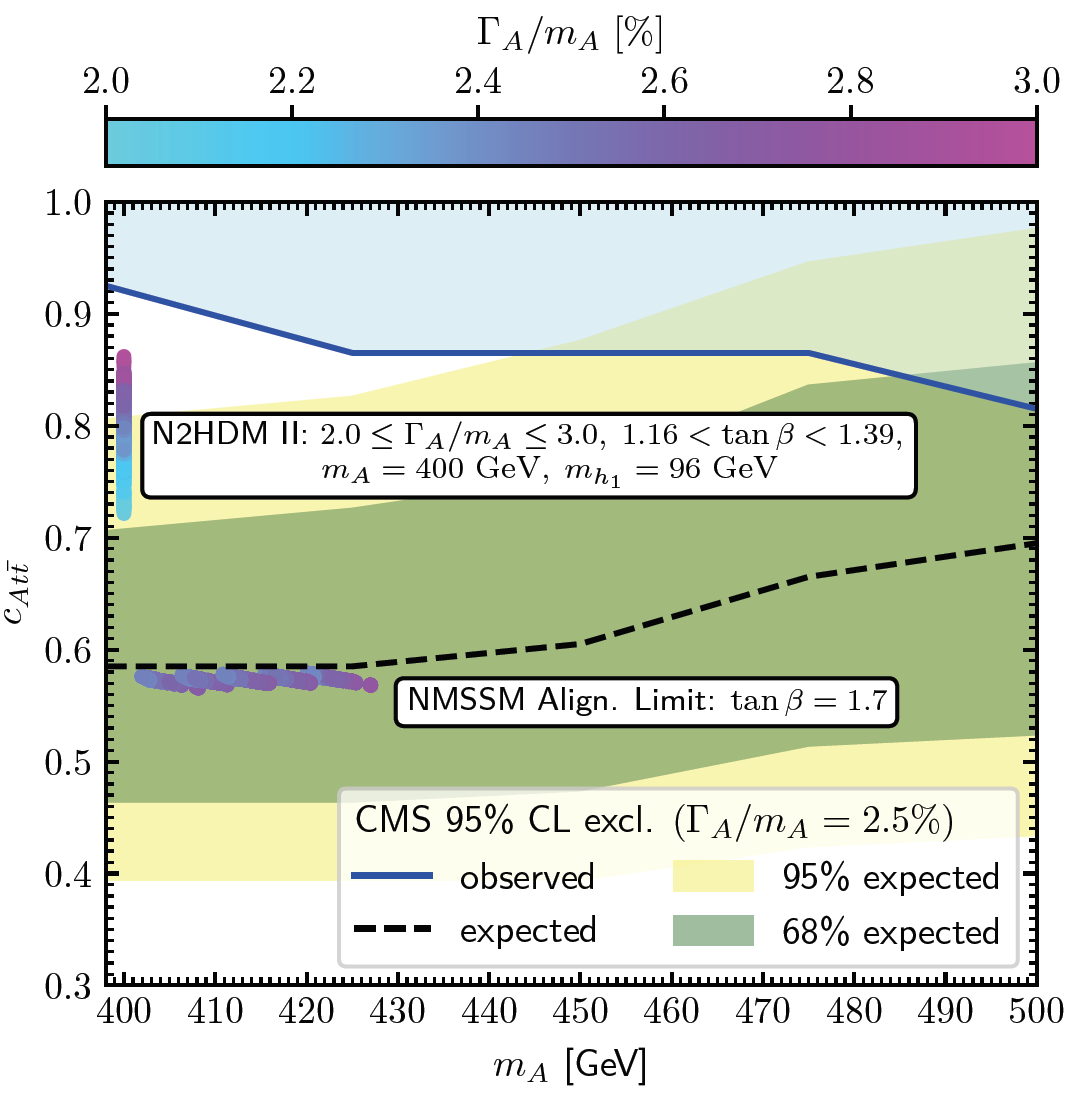}
  \includegraphics[width=0.45\textwidth]{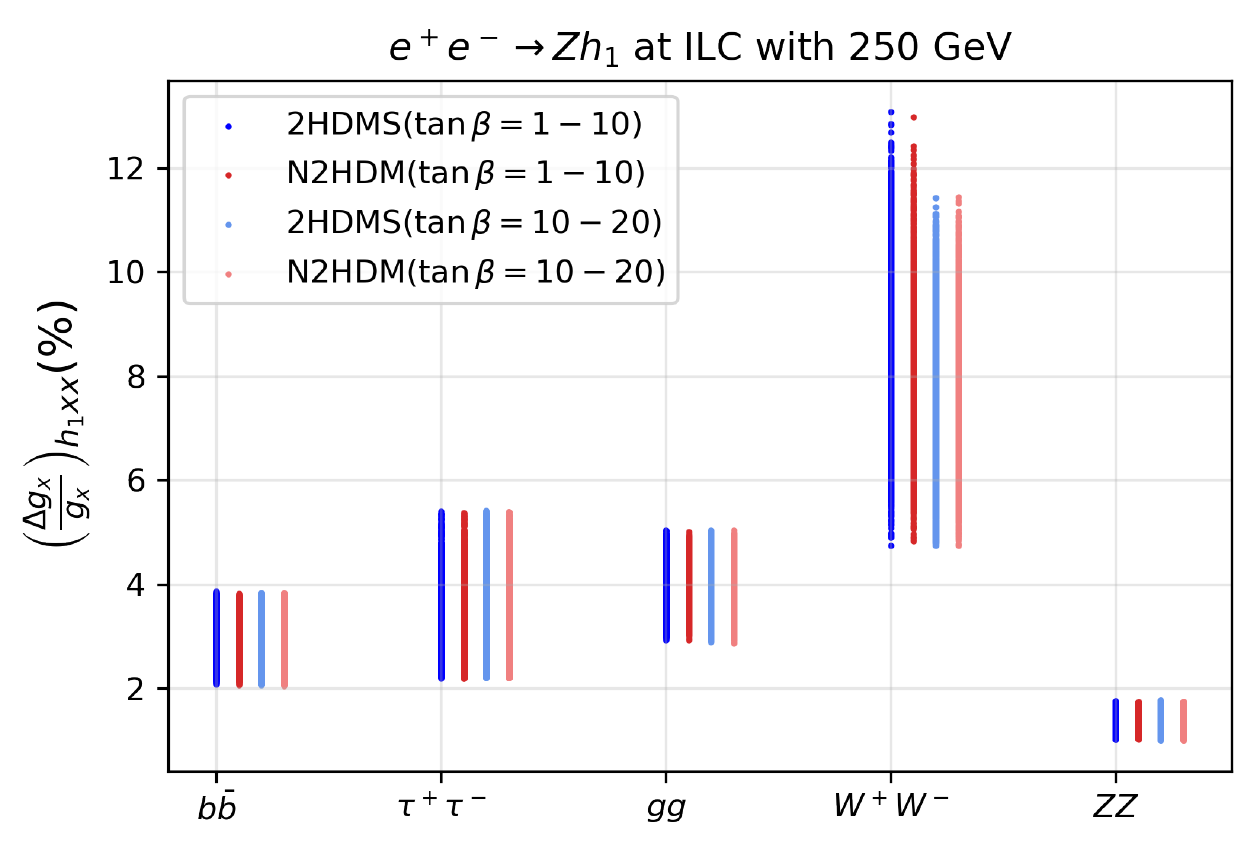}
  \caption{Left: $\MA$-$c_{At\bar t}$ plane in the N2HDM and NMSSM; color
    coding indicates $\Ga_A/\MA$ and where the $400 \gev$ and $96 \gev$ excesses
    can be realized~\cite{Biekotter:2021qbc} (see text).
  Right: anticipated coupling precision of $h_{96}$ in the N2HDM and
  2HDMS at the ILC250~\cite{Heinemeyer:2021msz} (see text).}
\label{fig:newhiggses}
\end{center}
\vspace{-2em}
\end{figure}

As a (toy) example, based on recent excesses observed in the Higgs
search data, we will briefly discuss one above and one below $125 \gev$.
The first one was seen by CMS in the $t\bar t$ final state at
$\sim 3\,\sig$ at a mass
scale of $\sim 400 \gev$~\cite{CMS:2019pzc}. It was best interpreted as
a $\cp$-odd Higgs boson produced in gluon fusion, $gg \to A \to t \bar t$, 
with $\Ga_A/\MA$ in the range of a few percent and a coupling strength
relative to an SM-like Higgs between $\sim 0.5$ and $\sim 0.9$. It was
shown in \citere{Biekotter:2021qbc} that this excess can be well
described by a $\cp$-odd Higgs boson in the Next-to-2HDM (N2HDM, where
the 2HDM is extended by a real singlet) or the Next-to-MSSM (NMSSM,
where the MSSM is extended by a complex singlet). Interestingly, the
indicated mass scale of $\sim 400 \gev$ would allow the production of
this particle at the ILC1000, or at CLIC1500 (or higher). The second
excess was seen at $\sim 96 \gev$ in the di-photon final state at
$\sim 3\,\sig$ at CMS~\cite{CMS:2018cyk} and in the $b\bar b$ final
state at $\sim 2\,\sig$ at
LEP~\cite{LEPWorkingGroupforHiggsbosonsearches:2003ing}. Such a state
could be produced (depending on its couplings to $Z$ bosons) in large
quantities at any future $e^+e^-$ collider operating at LEP energies, or
higher. In the left plot of \reffi{fig:newhiggses}~\cite{Biekotter:2021qbc}
the $\MA$-$c_{At\bar t}$ 
plane is shown, where $c_{At\bar t}$ is the coupling strength of the
$\cp$-odd Higgs to top quarks relative to an SM-like Higgs. The expected
exclusion line (black dashed) together with its 1~and $2\,\sig$
uncertainties, as well as the observed exclusion in light blue are taken
from \citere{CMS:2019pzc}. The colored points demonstrate where the N2HDM
can describe both excesses together, or where the NMSSM can describe the
$400 \gev$ excess (and possibly the di-photon excess at CMS, but not the
LEP excess), with the color coding indicating $\Ga_A/\MA$.

Interpreting the excesses in concrete models allows to make clear
predictions for current and future collider searches and
measurements. As discussed above, the \hnf\ can be produced at the first
stages of any of the future $e^+e^-$ colliders. Taking the ILC250 as a
concrete example, in \citere{Heinemeyer:2021msz} it was analyzed to
which precision the couplings of the \hnf\ can be measured. The
measurement of the $\hnf ZZ$ couplings proceeds directly via the
Higgs-Strahlung production mode, $e^+e^- \to Z\hnf$, with
$Z \to e^+e^-, \mu^+\mu^-$. The other coupling measurements rely on the
subsequent decay of the \hnf\ to the respective final state. As
underlying models the N2HDM as well as the 2HDMS (the 2HDM plus a complex
singlet), both of Yukawa type~II, were assumed. The anticipated
precisions in the various 
\hnf\ couplings can be seen in the right plot of
\reffi{fig:newhiggses}~\cite{Heinemeyer:2021msz}, where the color coding
distinguishes the two 
models, as well as two intervals for $\tb$. It can be observed that the
coupling to $Z$~bosons can be determined at the $1\%$~level, the
couplings to $b \bar b, \tau^+\tau^-, gg$ at the level of $2-5\%$, and
the coupling to $W^+W^-$ between $5\%$ and $12\%$. This demonstrates
that a future $e^+e^-$ collider can not only search and find new (light)
Higgs bosons, but can also perform a accurate analysis of its
couplings. This in turn will allow to determine the underlying paramter
space with high precision. 


\section{Theory meets Experiment}
\label{sec:theo-exp}

The discovery of a Higgs boson at the LHC required an enormous
experimental effort, and the LHC provided (and continues to provide) a
wealth of experimental data. However, the experimental data can only
fully be exploited if it is compared to theory predictions at the same
level of accuracy.%
\footnote{
The full uncertainty of a measured quantity is given by the (linear) sum
of the experimental and theoretical uncertainties.
}.%
~The non-background like events seen at the LHC in
the first $\sim 6\,\ifb$ of Run~1 data (plus the $7 \tev$ data) could be
interpreted as the discovery of a Higgs boson only by comparison to the
corresponding theory predictions. In order to provide these theory
predictions in the year 2010, i.e.\ two years prior to the Higgs-boson
discovery, the LHC Higgs Cross Section Working Group (now LHC Higgs
Working Group, LHCHWG) was founded~\cite{lhchwg}. Its task was (and is)
to provide theory predictions that match the (anticipated) experimental
accuracies in the search for Higgs bosons, as well as in the
interpretation of the (now discovered) signal. A related task was (and
is) to develop strategies for the extraction of experimental data on an
observed Higgs boson. As an example, the
\ka~framework~\cite{LHCHiggsCrossSectionWorkingGroup:2013rie} was
devised in 2012 to obtain information on the Higgs-boson couplings to SM
particles, see \refse{sec:hl-lhc}. The LHCHWG, composed jointly out of
theorists and experimentalists, continues to play a crucial role in the
interpreation of the observed signal at the LHC, as well as in the
search for new BSM Higgs bosons. 

The exploration of the \hotf\ at future $e^+e^-$
colliders will require an enormous improvement in the theoretical
predictions to meet the anticipated high experimental accuracy (see,
e.g., \citere{Heinemeyer:2006px}). The need for a coordinated effort to
match the expected experimental precision at an $e^+e^-$ collider is at
least as high, if not higher, as it was for the LHC. Therefore, the
theory effort should always be seen as an integral part of any (future)
Higgs (or SM/BSM) physics program.


\section{Complementarity}
\label{sec:compl}

One intriguing scenario for the future is the possible 
complementarity and synergy of collider based Higgs physics with
gravitional wave (GW) experiments. One prominent example for the realization
of BAU is baryogenesis, which requieres a strong first-order electroweak
phase transition (FOEWPT) in the early universe, which can take place
when the vacuum goes from a symetric phase to the broken phase: the
universe cools down and forms a second minimum in the Higgs potential
away from the origin. At the critical temperature $T_c$ both minima have
the same energy value and should be seperated by a barrier (otherwise a
second order phase transition or a smooth cross over can take
place). The universe then further cools down and at the nucleation
temperature $T_n$ it tunnels through the barrier from the symmetric phase to
the broken phase, the EW minimum with a vev of $\sim 246 \gev$. This in
turn generates GWs that could be detectable with GW observatories
such as the (approved) LISA~\cite{lisa1,lisa2}. 

\begin{figure}[htb!]
\begin{center}
  \includegraphics[width=0.55\textwidth]{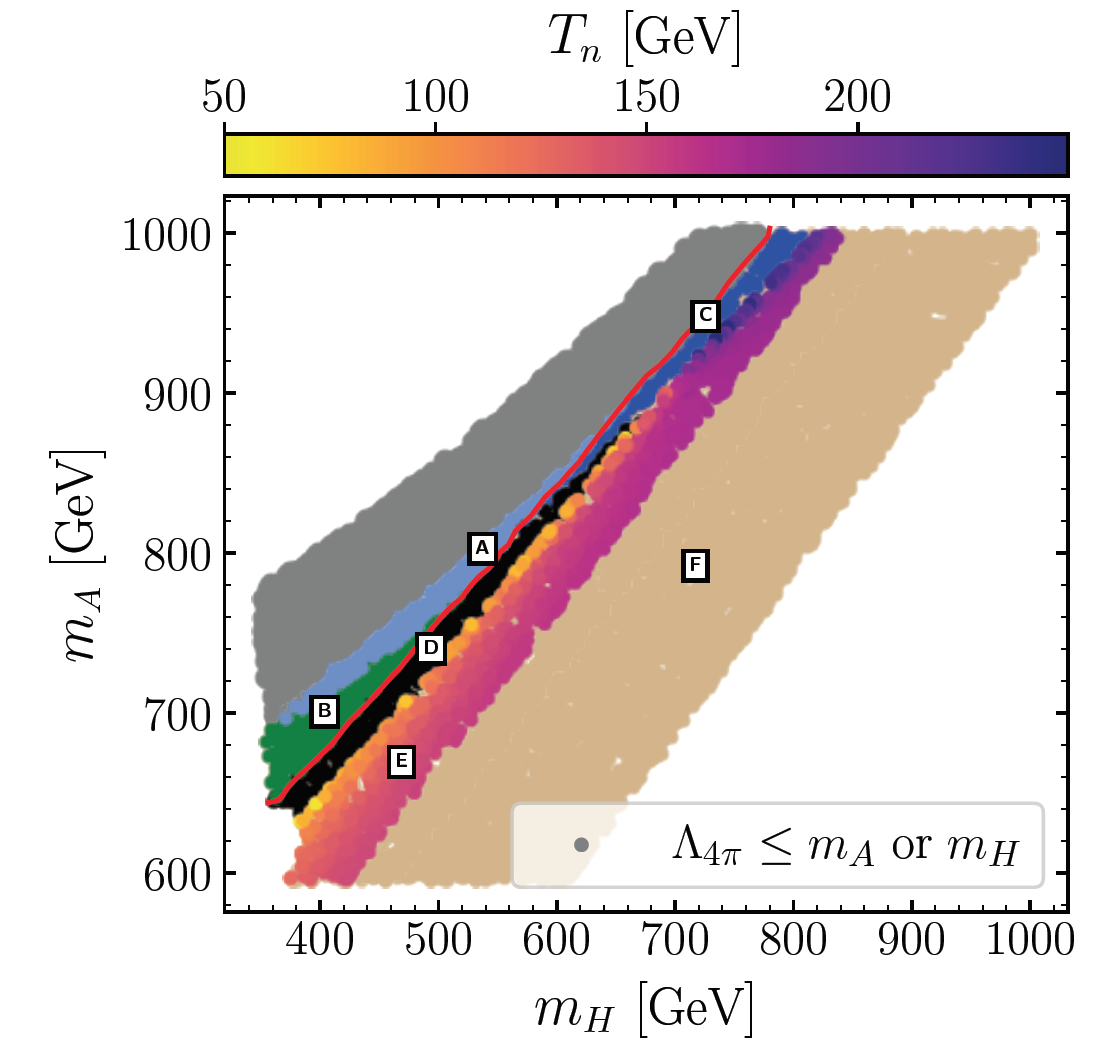}
  \caption{Thermal history of a 2HDM type~II scan~\cite{gw-2hdm} in the
    $\MH$-$\MA$ plane. The color coding indicates the nucleation
    temperature $T_n$ (in zone~E, see text).}
\label{fig:GW-2HDM-FOEWPT}
\end{center}
\vspace{-2em}
\end{figure}

In the SM this is not
possible with the measured Higgs-boson mass of $\sim 125 \gev$. In BSM
Higgs sectors with additional degrees of freedom, however, this
constitutes a realistic possibility. As an example we show in
\reffi{fig:GW-2HDM-FOEWPT}~\cite{gw-2hdm} a parameter plane in the 2HDM
type~II with $\Mh = 125 \gev$, $\tb = 3$, $\cos(\be-\al) = 0$,
$m_{12}^2 = \MH^2 \sin\be \, \cos\be$, and with $\MA = \MHp$ and $\MH$ as
free parameters. In the benchmark plane various zones (A-F) can be
distringuished by their thermal history of the universe. In zone~E the
above discussed FOEWPT takes place, with the color coding indicating
$T_n$, where lower $T_n$ yields a stronger GW signal. Here it is
interesting to observe that the parameter space with the potentially
largest GW signal, located in zone~D shown in black, features ``vacuum
trapping'': the 
broken phase is the deepest minimum, but the barrier between the
symmetric phase and the EW minimum is so strong that no transition takes
place. On the other hand, if zone~E is realized in nature and the GW
signal is sufficiently strong (to be detected by LISA or other possible
future GW observatories) the collider experiments for BSM Higgs sectors
and the GW detectors yield complementary information about the BSM Higgs
sector. 

\begin{figure}[htb!]
\begin{center}
  \includegraphics[width=0.85\textwidth]{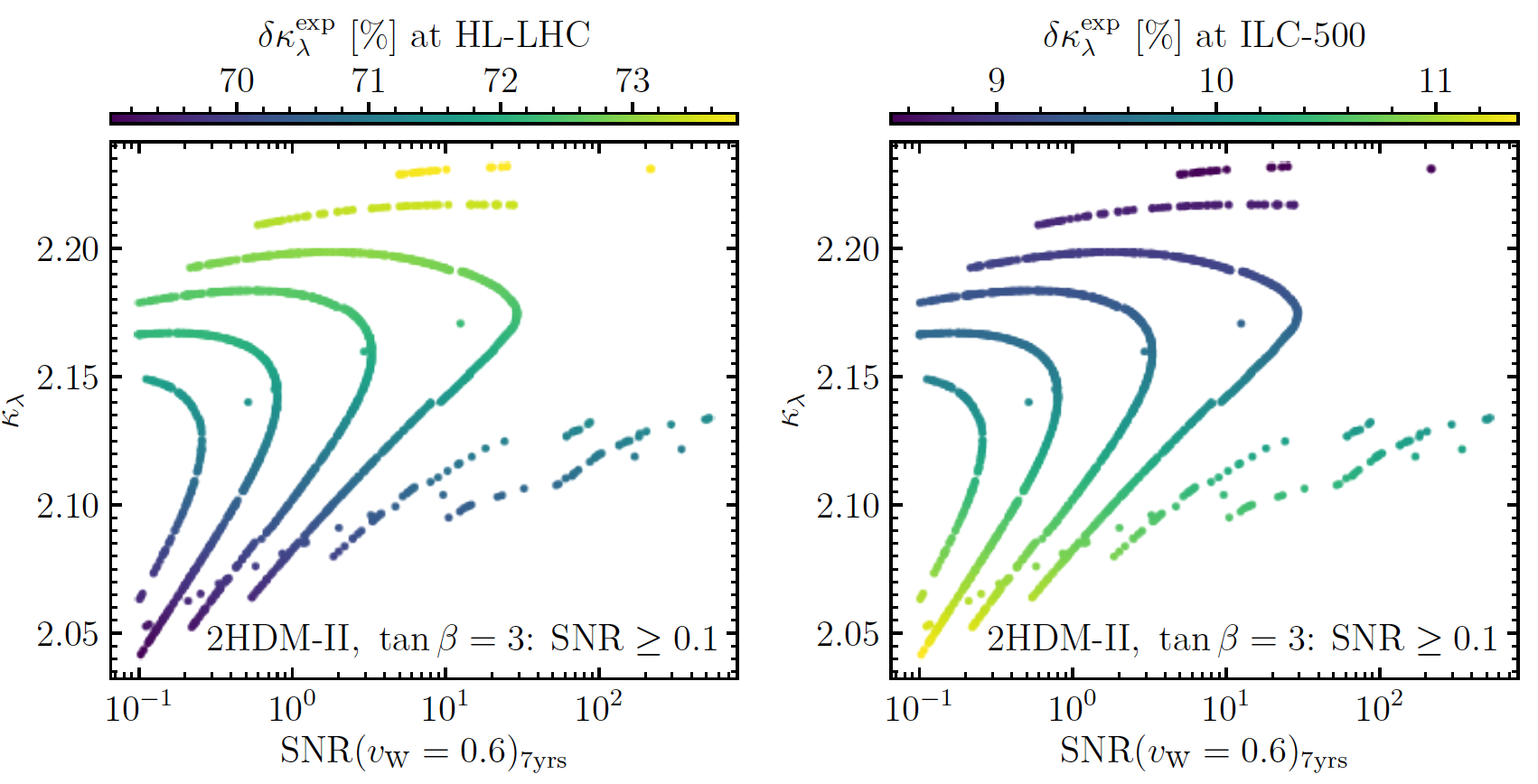}
  \caption{Parameter points from a 2HDM type~II scan~\cite{gw-2hdm} in
    the LISA SNR-\kala\ plane (see text). The color code indicates the
    precision $\de\kala^{\rm exp}$ at the HL-LHC (left) and the ILC500
    (right).} 
\label{fig:GW-2HDM-lahhh-prec}
\end{center}
\vspace{-2em}
\end{figure}

As a final step in this analysis we show in
\reffi{fig:GW-2HDM-lahhh-prec}~\cite{gw-2hdm} the parameter space of the
previous benchmark plane that yields a signal-to-noise ratio (SNR) at
LISA larger than 0.1, as shown in the horizontal axis
(using 7~years of data taking and a bubble wall velocity of
$v_{\mathrm W} = 0.6$).
An observable signal will likely require an SNR
larger than~1. It is found that in this area the higher-order
corrections to \lahhh\ yield values of $\kala \sim 2$, i.e.\ a strong
enhancement w.r.t.\ the SM value, as can be seen in the vertical axis.
This is a typical result for FOEWPT in the 2HDM.
The color coding in the figure indicates the precision with which the
HL-LHC (left plot) and the ILC500 (right plot) can determine
$\kala$. The anticipated precision changes because of the changed
interferences of signal and background diagrams contributing to the
di-Higgs production cross sections.
One can see in \reffi{fig:GW-2HDM-lahhh-prec} that the HL-LHC precision
worsened substantially w.r.t.\ the SM, i.e.\ $\kala = 1$, down to $\sim 72\%$. 
On the other hand, the precision at the ILC500 improves by about a
factor of $\sim 2$ w.r.t.\ $\kala = 1$, yielding a determination at the
level of $\sim 10\%$. The important overall conclusion is that if BAU is
generated by baryogenesis, requiering a 
FOEWPT in the early universe, this likely worsens the prospects to
determine \lahhh\ at the HL-LHC, but substantially improves the ILC500
prospects. 


\section{Main conclusions}
\label{sec:concl}

After the full exploitation of the HL-LHC Higgs data, 
let's build an $e^+e^-$ collider as quickly as possible to study the
observed 
Higgs boson in detail and to search for new Higgs bosons above and below
$125 \gev$.


\end{document}